# On Uses of Mean Absolute Deviation: Shape Exploring and Distribution Function Estimation

**Elsayed A. H. Elamir**
Department of Management and Marketing,
College of Business Administration, Kingdom of Bahrain
Email: shabib@uob.edu.bh

**Abstract**

Mean absolute deviation function is used to explore the pattern and the distribution of the data graphically to enable analysts gaining greater understanding of raw data and to foster quick and a deep understanding of the data as an important fundament for successful data analytic. Furthermore, new nonparametric approaches for estimating the cumulative distribution function based on the mean absolute deviation function are proposed. These new approaches are meant to be a general nonparametric class that includes the empirical distribution function as a special case. Simulation study reveals that the Richardson extrapolation approach has a major improvement in terms of average squared errors over the classical empirical estimators and has comparable results with smooth approaches such as cubic spline and constrained linear spline for practically small samples. The properties of the proposed estimators are studied. Moreover, the Richardson approach applied for real data application and used to estimate the hazardous concentration five percent.

**Keywords –** Empirical distribution function, Nonparametric estimation, numerical differentiation, Richardson extrapolation, Skewness, Uniform consistency

Subject classification code: 62G30 and 62G32

# 1 Introduction

Data analytic techniques are very important to explore the structure and the distribution of the data to enable data analysts gaining greater understanding of the raw data. Data is often collected in large, unstructured volumes from various sources and analysts must first understand and develop a comprehensive view of the data before extracting relevant data for further analysis; see, Healy (2019). Nowadays, data exploration has established as a mandatory phase in every data science project. Typical plots include scatter plot, histograms, box plot, quantile quantile plot and many more have been used as a graphical approach to learn about distributions, correlations, outliers, trends, and other data characteristics; see, Tukey (1987), Gandomi and Haider (2015) and Cumming and Finch (2005). One main advantage of data exploration graphically is to learn about characteristics and potential problems of a data set without the need to formulate assumptions about the data beforehand and to foster quick and a deep understanding of the data as an important fundament for successful and efficient data science projects; see, Matt and Joshua (2019), Runkler (2020), James et al. (2013), Healy (2019) and Larson-Hall (2017).

The estimation of a distribution function is not only a fascinating problem by itself, but it also emerges naturally in real-world problems in a variety of scientific domains, including commerce, hydrology, and environmental sciences. As a result, a variety of nonparametric approaches for tackling this problem have risen in different disciplines; see, Efromovich (2001), Cheng and Peng (2002), and Charles et al. (2010). The risk term, or natural hazard, appears to be closely tied to the distribution function in many circumstances. Scientists want to know the likelihood of a large earthquake, the likelihood of high wind speeds or hurricanes, and the hazard of low levels; see, Baszczynska (2016), Xue and Wang (2009) and Babu et al. (2002) and Erdogan et al. (2019).

The population mean absolute deviation (MAD) about any value $v$ can be written as

$$\Delta_X(v) = E|X - v|, \qquad v \in R \tag{1}$$

This function is usually used as a direct measure of the scale for any distribution about chosen $v$ such as mean absolute deviation about population mean ($\mu$)

$$\Delta_X(\mu) = E|X - \mu|$$

and mean absolute deviation about population median ($M$)

$$\Delta_X(M) = E|X - M|$$

These measures offer a direct measure of the dispersion of a random variable about its mean and median, respectively, and have many applications in different fields; see, Dodge (1987), Pham-Gia and Hung (2001), Gorard (2005), Elamir (2012) and Habib (2012).

This article is organized as follows. MAD function representation is explained in Section 2. MAD plot is proposed in Section 3. Some uses of MAD function are introduced in Section 4. Distribution function in terms of MAD is presented in Section 5. Several nonparametric estimation approaches for distribution function are derived in Section 6. Simulation study is conducted to study the properties of proposed estimators in terms of average mean square in Section 7. Ricardson extrapolation approximation is applied to acute toxicity values in Section 8. Section 9 is devoted for conclusion.

# 2 MAD function and its representations

Let $X_1, \dots, X_n$ be an independent and identically a random sample from a continuous distribution function $F_X(.)(0 < F < 1)$, density $f_X(.)(f \geq 0)$, mean $\mu = E(X)$, median $M =$ $Med(X)$, standard deviation $\sigma = \sqrt{E(x-\mu)^2}$, indicator function $I_{i\leq k}$ is 1 if $i \leq k$, 0 else, and $X_{(1)}, \ldots, X_{(n)}$ the corresponding order statistics.

The population mean absolute deviation (MAD) about any value $v$ can be written as

$$\Delta_X(v) = E|X - v|, \quad v \in R \tag{2}$$

Another representation of MAD in terms of distribution function is given by Munoz-Perez and Sanchez-Gomez (1990) as

$$\Delta_X(v) = v[2F_X(v) - 1] + E(X) - 2E[XI_{X \leq v}] \tag{3}$$

and its first derivative

$$\acute{\Delta}_X(v) = 2F_X(v) - 1 \tag{4}$$

For more details, see, Habib (2012).

**Theorem 1.** The mean absolute deviation about $v$ ($v \in R$) is minimized when $v$ is the median and it is a convex function.

*Proof.*

Since $F_X(M) = 0.5$, the first derivative of MAD function at median $(M)$ is zero ($\acute{\Delta}_X(M) = 0$) with positive second derivative $\acute{\acute{\Delta}}_X(M) = 2f_X(M) > 0$. Therefore, $\Delta_X(v) = E|X - v|$ has a minimum value at $v = M$. Where $\acute{\Delta}_X(v) = 2F_X(v) - 1$ and $\acute{\acute{\Delta}}_X(v) = 2f_X(v) \geq 0$ for all $v \in R$, then $\Delta_X(v)$ is a convex function.

Munoz-Perez and Sanchez-Gomez (1990) proved that $\Delta_X(v)$ characterizes the distribution function and give a dispersive ordering of probability distributions as it satisfies the following conditions: (1) There are only a finite number of discontinuity points in the derivative, (2) it is a convex function on real line $(R)$, (3) the $\lim_{x \to \infty} \acute{\Delta}_X(x) = 1$ and $\lim_{x \to -\infty} \acute{\Delta}_X(x) = -1$ and (4) the $\lim_{x \to \infty}[\Delta_X(x) - x] = -E(X)$, and $\lim_{x \to -\infty}[\Delta_X(x) + x] = E(X)$. Since $\Delta_X(v)$ satisfies the above conditions, there subsists a unique distribution function which has $\Delta_X(v)$ as its dispersion function.

## 3 Mean absolute deviation plot (MAD plot)

The MAD plot can be introduced as

$$X_{axis} = v_i, \quad \text{versus} \quad Y_{axis} = \Delta_X(v_i), \text{ for each } v_i = x_i \text{ and } i = 1, \ldots, n$$

with two straight lines at

$$\mu - v_i \text{ and } v_i - \mu$$

This plot represents data on x-axis and the mean absolute deviation at each $v = x$ on the y-axis that includes mean absolute deviation about mean and median as special cases. In other words, it is a simple curve plot between actual data and its mean absolute deviation at each point. Figure 1 displays the MAD plot for standard normal distribution using the quantile function for standard norm from R-software (2021) $v_i = qnorm(p = (i - 0.5)/n, \mu = 0, \sigma = 1)$, and $i = 1, \ldots, 100$ with two straight lines $\mu - v_i$ and $v_i - \mu$ that shows the degree of approximation with $\Delta_X(v_i)$.

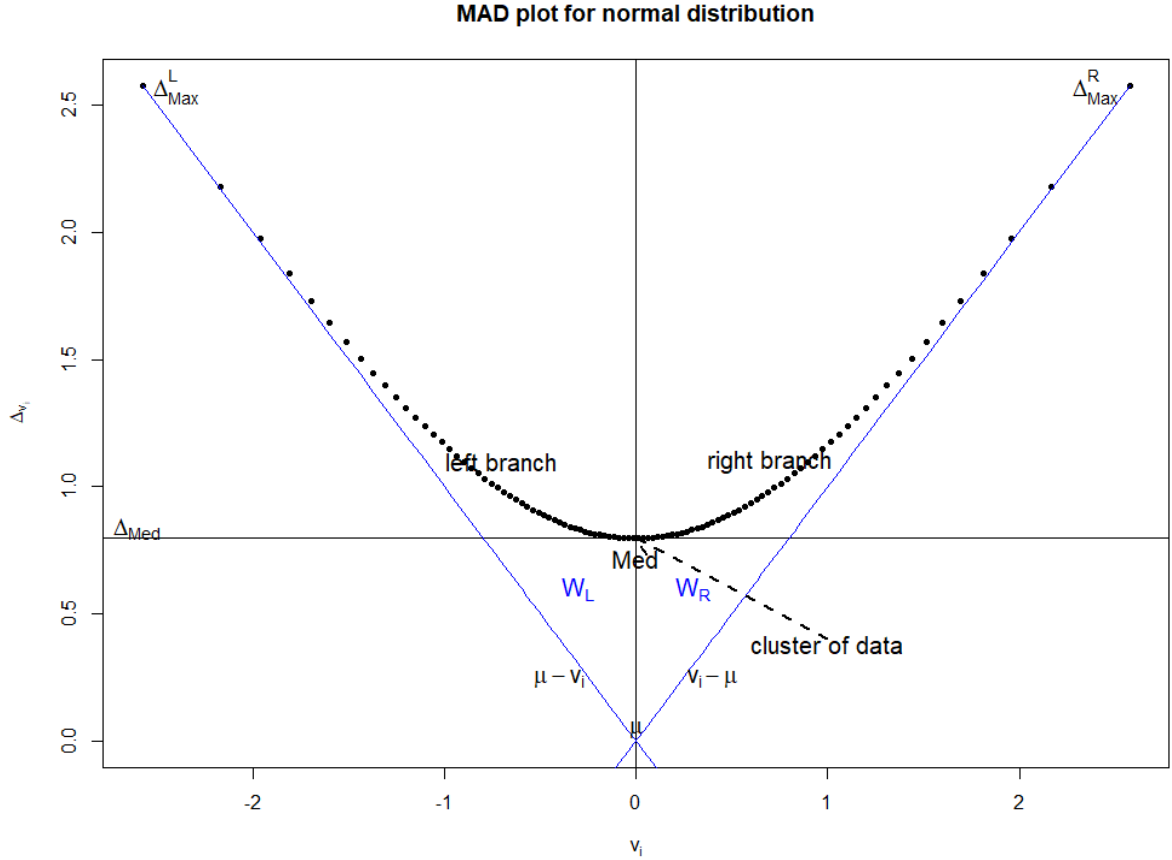

Figure 1 MAD function plot for standard normal using $qnorm((i - 0.5)/n)$, and $i = 1, \ldots, 100$.

For the normal distribution, the MAD function formulates a parabola curve, or a quadratic function that has a minimum at median $\Delta_X(M) = \Delta_{Med}$ (mean absolute deviation about median) and reflects a lot of information that includes

- The location measures median ($M$) on x-axis at $min\ \Delta_X(v)$ and mean ($\mu$) at the intersection of straight lines,
- the scale measure $\Delta_X(v)$ (mean absolute deviation about median) at minimum of the MAD function,
- the right MAD branch $\Delta_X(v)I_{v>M}$ and its maximum $\Delta_{Max}^R$ that gives an indication of spread out of the data and tail length in the right side,
- the left MAD branch $\Delta_X(v)I_{v<M}$ and its maximum $\Delta_{Max}^L$, that gives an indication of spread out of the data and tail length in the left side,
- two straight lines $\mu - v_i$ and $v_i - \mu$ with intersection at $\mu$ that gives the degree of approximation with $\Delta_X(v_i)$,
- The wideness between MAD function, median, and straight lines $\mu - v_i$ and $v_i - \mu$ (right wideness, $W_R$, and left wideness $W_L$) that give an indication of the direction and degree of skewness and peakedness,
- The cluster of data that may give an indication about modality in the data.

Furthermore, the MAD function could be divided to the right MAD function ($\Delta_X^+(v)$) and the left MAD function ($\Delta_X^-(v)$) as

$$\Delta_X(v) = E|X - v| = E(X - v)^+ + E(X - v)^- = \Delta_X^+(v) + \Delta_X^-(v)$$

and

$$\Delta_X^+(v) = E[(X-v)I_{X>v}] \quad \text{and} \quad \Delta_X^-(v) = E[(v-X)I_{X\leq v}]$$

The relationship between the straight lines and the MAD functions can be written as

$$E(X-v) = \Delta_X^+(v) - \Delta_X^-(v)$$

Therefore, when $v = \mu$, we have $\Delta_X^+(v) = \Delta_X^-(v)$, also if $v = M$, we have $\mu - M = \Delta_X^+(M) - \Delta_X^-(M)$ that could be considered as measure of skewness; see, Habib (2012) and Elamir (2012).

## 4 Uses of MAD functions

### 4.1 Wideness and skewness

The area between right straight line, median and right MAD branch can be defined in standard form as

$$W_R = \frac{\Delta_X^+(M)}{\sigma}$$

and

$$\Delta_X^+(M) = E[(X-M)I_{X>M}]$$

It may consider as the right wideness measure which reflect how much the right MAD branch is away from the right straight line $(v_i - \mu)$ and median. Since $\Delta \leq \sigma$ by Jensen's inequality, it is straightforward to prove that $0 \leq W_R \leq 1$. In terms of data, when the value of $W_R$ is near 1 it indicates big wideness or stretch out from the median, in other words, the data will be spread out far away from median in the right side. If value of $W_R$ is near 0 it indicates small wideness from the median, in other words, the data will be concentrated near the median in the right side.

Also, the area between the left straight line, median and the left MAD branch (left wideness) can be defined in standard form as

$$W_L = \frac{\Delta_X^-(M)}{\sigma}$$

and

$$\Delta_X^-(M) = E[(M-X)I_{X\leq M}]$$

It may consider as the left wideness measure which reflect how much the left MAD branch is away from left straight line $(\mu - v_i)$ and median. It is straightforward to prove that $0 \leq W_L \leq 1$. In terms of data, when the value of $W_L$ is near 1 it indicates big wideness or stretches out from the median. In other words, the data will be spread out far away from median in the left side. If value of $W_L$ is near 0 it indicates small wideness from the median, in other words, the data will be closed to the median in the left side.

The general measure of wideness for a distribution in terms of right and left wideness may be defines as

$$W = W_L + W_R = \frac{\Delta_X^-(M) + \Delta_X^+(M)}{\sigma} = \frac{\Delta_X(M)}{\sigma}$$

It may be considered as a measure of total wideness between MAD function and the two straight lines. This measure will be very useful for symmetric distributions where it may be related to what is called platykurtic or flatness that had been used as a test for normal distribution; see, Geary (1935) and Elamir (2012). The interpretation of this measure especially for symmetric distributions in terms of data can be as follows. If $W$ is near 1, the distribution of the data is strong "curved inwards", near zero strong "curved outwards", and 0.798 normal like in terms of wideness.

The tightness between MAD function and the two straight lines may be defined as a complement of wideness as

$$L = 1 - W = \frac{\sigma - \Delta_X(M)}{\sigma}$$

The standardized distance between the standard deviation of the population and the mean absolute deviation about median. This measure is very useful for symmetric distributions where it may be related to what is called leptokurtic (peakedness). As $\sigma$ getting far away from $\Delta_X(M)$, the more leptokurtic (more data concentration about median). Since $\Delta_X(M) \leq \sigma$, then $0 \leq L \leq 1$. The values of $W_R$, $W_L$, $W$ and $L$ are presented in Table 1 for some selected symmetric distributions. The distributions that have big flatness as Beta (1, 1) and Beta (0.1, 0.1) have $W$ near 1 while the distributions with strong peakedness such as $t(df = 3)$ has less value.

Table 1 Wideness and leptokurtic for some symmetric distributions

| Distribution | $W_R$ | $W_L$ | $W$ | $L$ |
|---|---|---|---|---|
| Beta(0.1,0.1) | 0.483 | 0.483 | 0.966 | 0.034 |
| Beta(1,1) | 0.433 | 0.433 | 0.866 | 0.134 |
| Normal | 0.399 | 0.399 | 0.798 | 0.202 |
| Logistic | 0.382 | 0.382 | 0.765 | 0.235 |
| Laplace | 0.355 | 0.355 | 0.710 | 0.290 |
| $t(df = 3)$ | 0.326 | 0.326 | 0.652 | 0.348 |

With respect to the tails of data distribution, different measures may be proposed from Figure 1 as

$$T_R = \frac{\Delta_{Max}^R}{\sigma} \quad \text{and} \quad T_L = \frac{\Delta_{Max}^L}{\sigma}$$

Alternatively,

$$T_{R1} = \frac{\sigma}{\Delta_{Max}^R} \quad \text{and} \quad T_{L1} = \frac{\sigma}{\Delta_{Max}^L}$$

where

$$\Delta_{Max}^R = Max.[(v - X)I_{X \leq v}] \quad \text{and} \quad \Delta_{Max}^L = Max.[(X - v)I_{X > v}]$$

All values of these measures are more than or equal 0 with no upper value. These measures give an indication about tail length. For the first two measures, the small values around 1 indicates short tails, around 3 medium tails while larger value indicate heavy (long) tail. For the other two measures $T_{R1}$ and $T_{L1}$, the values around zero very heavy (long) tails, values around 0.30 medium tails while values around one light (short) tails. Note that the values of the above measures will depend on sample size.

The first measure of skewness in terms of wideness can be defined as

$$SK_1 = W_R - W_L = \frac{\mu - M}{\sigma}$$

This measure is equivalent to Groeneveld and Meeden (1984) measure of skewness. The second measure of skewness in terms of tailedness can be proposed as

$$SK_2 = T_R - T_L = \frac{\Delta_{Max}^R - \Delta_{Max}^L}{\sigma}$$

Alternatively,

$$SK_{21} = T_{L1} - T_{R1} = \frac{\sigma}{\Delta_{Max}^L} - \frac{\sigma}{\Delta_{Max}^R}$$

Figure 2 displays MAD plot for Beta (0.1,0.1), normal, Laplace and exponential distributions using quantile function for each distribution in R-software with $n = 300$. It may conclude that

- the location measures median and mean values are located at minimum of MAD curve and intersection between two straight lines on x axis, respectively, while the dispersion measure $\Delta_X(M)$ is located opposite of min. MAD curve on y axis,
- tails measures ($T_R$ and $T_L$) around 1 may give an indication of short tail such as beta while around 3 may give an indication normal alike,
- equal wideness ($W_R, W_L$) may give an indication of symmetric distribution such as beta, normal and Laplace, while not equal measures are indication of skewed distributions such as exponential,
- the data cluster at both ends of the curve may give an indication for bimodality distribution such as beta (0.1, 0.1) and at one position may give an indication of unimodal distributions such as, normal, Laplace and exponential,
- for symmetric distributions, value for wideness as $0.483 + 0.483 = 0.966$ may give an indication of strong "curved inwards" such as beta (0.1,0.1) and value as 0.355+0.355=0.71 may give an indication of data close to median and peakedness,
- equal tail lengths may give an indication of symmetric distribution such as beta, normal and Laplace while not equal gives and indication of skewed distributions, such as exponential.

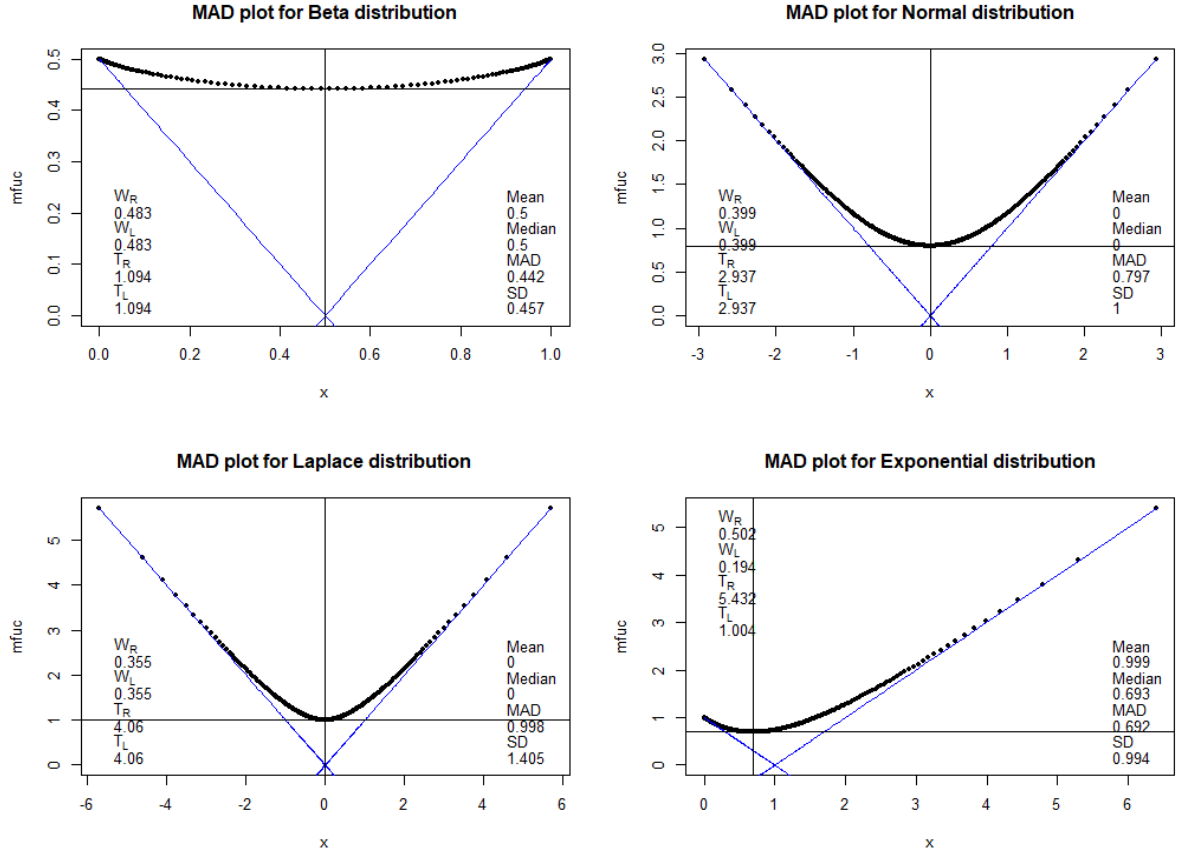

Figure 2 MAD plot for Beta (0.1,0.1), normal, Laplace and exponential distributions using quantile function for each distribution in R-software with $p = (i - 0.5)/n$, and $i = 1 \ldots 300$.

## 5 Distribution function in terms of MAD

According to Munoz-Perez and Sanchez-Gomez (1990), the MAD function can be rewritten in terms of indicator function as

$$\Delta_X(v) = E[(X - v)I_{X>v}] + E[(v - X)I_{X \leq v}]$$

The first derivative of $\Delta_X(v)$ with respect to $v$ can be obtained as

$$\acute{\Delta}_X(v) = -[1 - F_X(v)] + F_X(v)$$

Therefore, the distribution function can be re-expressed in three MAD functions as follows. The distribution function can be rewritten in terms of derivative of MAD function ($\acute{\Delta}_X(v)$) as

$$F_X(v) = \frac{\acute{\Delta}_X(v) + 1}{2} = 0.5 + 0.5\acute{\Delta}_X(v) \qquad (5)$$

In terms of derivative of right MAD function ($\acute{\Delta}_X^+(v)$) as

$$F_X(v) = 1 + \acute{\Delta}_X^+(v) \qquad (6)$$

Finally, in terms of derivative of left MAD function as

$$F_X(v) = \acute{\Delta}_X^-(v) \qquad (7)$$

The right and left MAD can characterize the distribution function because they are primeval function of $F_x$; see, Munoz-Perez and Sanchez-Gomez (1990).

**Theorem 2.** The right and left MAD functions $\Delta_X^+(v)$ and $\Delta_X^-(v)$ are monotone decreasing and increasing functions, respectively.
*Proof:*
Where $\acute{\Delta}_X^+(v) = F_X(v) - 1 \le 0$ for all $v \in R$, $\Delta_X^+(v)$ is a monotone decreasing function. Since $\acute{\Delta}_X^-(v) = F_X(v) \ge 0$ for all $v \in R$, $\Delta_X^-(v)$ is a monotone increasing function.

Figure 3 displays the MAD functions ($\Delta_X(v), \Delta_X^+(v)$ and $\Delta_X^-(v)$) plot of standardized data from beta, exponential, Laplace and normal distributions. It may conclude that
- $\acute{\Delta}_X^-(v)$ is monotone increasing function with minimum 0 and intersection with $\Delta_X^+(v)$ at mean,
- $\acute{\Delta}_X^+(v)$ is monotone decreasing function with minimum 0 and intersection with $\Delta_X^-(v)$ at mean,
- $\acute{\Delta}_X^-(v)$ and $\acute{\Delta}_X^+(v)$ have joint points with $\Delta_X(v)$ at extreme ends of the $\Delta_X(v)$ curve.

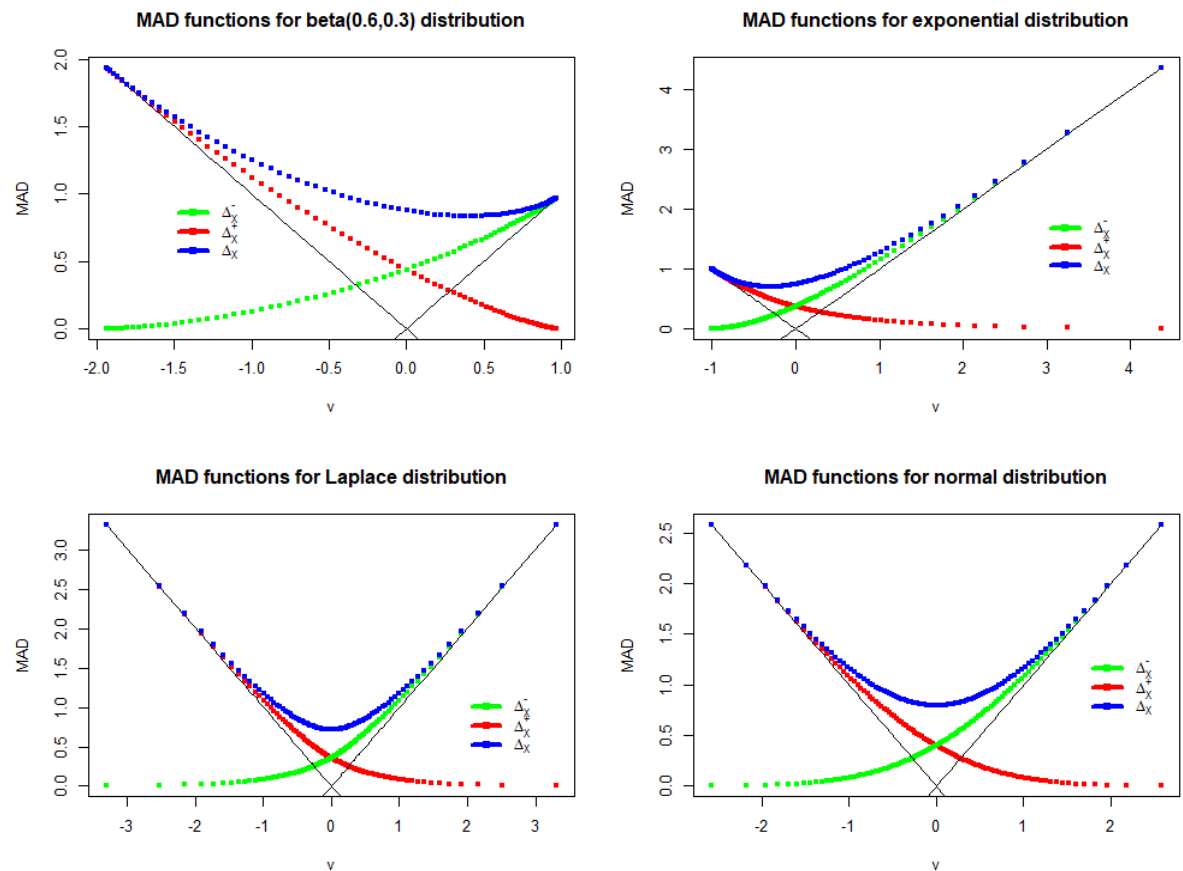

Figure 3 MAD functions ($\Delta_X(v), \Delta_X^+(v)$ and $\Delta_X^-(v)$) plot of standardized data from beta (0.6, 0.3), exponential, Laplace and normal distributions using quantile functions with $q = (i - 0.5)/n$ and $n = 100$

The most common non-parametric estimator for the underlying distribution function $F$ is specified by the empirical cumulative distribution function (ecdf). The ecdf is defined by

$$\hat{F}_n(x) = \frac{1}{n}\sum_{i=1}^{n} I_{\{X_i \leq x\}}$$

$\hat{F}_n(x)$ has good statistical properties such as: (1) it is a nondecreasing function with jumps of size $1/n$ at each order statistic, (2) it is bounded between zero and one, (3) it is first order efficient based on minimax criteria, for every $x$, $h = n\hat{F}_n(x)$ has a binomial distribution $(n, p = F_X(x))$, (4) and for large $n$, $\sqrt{n}(\hat{F}_n(x) - F(x)) \sim N(0, F_X(x)(1 - F_X(x)))$; see Dvoretzky et al. (1956), Lehmann and Cassella (1998), and Csaki (1984). Furthermore, the empirical distribution function is the nonparametric maximum likelihood estimator of $F$ and has an important role in nonparametric bootstrap and simulation; see, Haddou and Perron (2006) and Efron and Tibshirani (1993).

## 6  Estimation of distribution function using MAD

When it is difficult to analytically obtain the derivative of the function $\Delta_X^-(v)$, the derivative of the function $\Delta_X^-(v)$ at $v$ is

$$F_X(v) = \lim_{h \to 0} \frac{\Delta_X^-(v+h) - \Delta_X^-(v)}{h} \tag{8}$$

The numerical derivative can be used to obtain a good approximation to the true function $F_X(v)$. In the following we assume that this limit exists, i.e., $\Delta_X^-(v)$ is differentiable at $x = v$. By using the numerical differentiation, it could consider nonparametric estimators of the population distribution function $F_X(v)$ using a random sample $X_1, \ldots, X_n$ of size $n$. Consider the pairs of data

$$(x_i = x_{(i)}, y_i, i = 1, \ldots, n)$$

where $x_{(i)}$ is the observed order data and $y_i$ is the estimated left MAD function that can be obtained from data as

$$y_i = g(x_{(i)}) = \widehat{\Delta_X^-(v)} = \hat{E}[(v-x)I_{x \leq v}] =$$
$$= \frac{1}{n}\sum_{j=1}^{n}(v_i - x_j)I_{x_j \leq v_i}, \text{ for each } v_i = x_{(i)}, \quad i = 1, \ldots, n$$

The nonparametric estimates of $F_X(v)$ can be derived numerically using several approaches as follows.

### 6.1  Forward difference approach

By using Taylor series

$$g(x+h) = g(x) + \acute{g}(x)h + \frac{\acute{g}(x)}{2}h^2 + \cdots + \frac{g^{k-1}(x)}{(k-1)!}h^{k-1}$$

See; Burden and Faires (2011) and Levy (2012).
The first derivative in terms of first two terms,

$$\acute{g}(x) \sim \frac{g(x+h) - g(x)}{h} - \frac{h^2}{2}\acute{g}(\xi), \quad \xi \in (x, x+h)$$

The first order approximation $O(h)$. Therefore, a forward difference approach is

$$\acute{g}(x_i) \sim \frac{g_{i+1} - g_i}{h} + O(h)$$

By considering $h = x_{i+1} - x_i$ and with one-sided 1 at the endpoints of the data set, an estimation of $F_X(v)$ can be approximated by two terms Taylor series expansion as

$$\hat{F}_O(v) = \begin{cases} \frac{y_{i+1} - y_i}{x_{(i+1)} - x_{(i)}}, & i = 1, \dots, n-1 \\ 1, & i = n \end{cases} \quad (9)$$

**Theorem 3.** with one-sided 1 at the endpoints of the data set, the forward difference approach is

$$\hat{F}_O(v) = \begin{cases} \frac{i}{n}, & i = 1, \dots, n-1 \\ 1, & i = n \end{cases}$$

*Proof:* by considering $y_i = \frac{1}{n}\sum_{j=1}^n (v_i - x_j) I_{x_j \leq v_i}$, for each $v_i = x_i$, $i = 1, \dots, n$.
For $v_1 = x_1$ then $y_1 = 0$, for $v_2 = x_2$ then $y_2 = \frac{1}{n}(x_2 - x_1)$, for $v_3 = x_3$ then $y_3 = \frac{1}{n}[(x_3 - x_1) + (x_3 - x_2)] = \frac{1}{n}[2x_3 + ny_2 - x_2 - x_2] = \frac{1}{n}[2(x_3 - x_2) + ny_2]$, ...,
for $v_i = x_i$ then $y_i = \frac{1}{n}[(x_i - x_1) + \dots + (x_i - x_{i-1})] = \frac{1}{n}[(i-1)(x_i - x_{i-1}) + ny_{i-1}]$,
therefore,

$$y_i - y_{i-1} = \frac{(i-1)}{n}(x_i - x_{i-1}), \quad i = 2, \dots, n$$

and

$$y_{i+1} - y_i = \frac{i}{n}(x_{i+1} - x_i), \quad i = 1, \dots, n-1$$

This shows that the forward difference approach for left MAD function is just the empirical distribution function $i/n, i = 1, \dots, n$ and has equal jumping value $1/n$.

## 6.2 Backward difference approach

Similarly, a backward first order approximation $O(h)$ or a backward differencing estimation of $F_X(v)$ can be approximated by two terms Taylor series expansion as

$$\hat{F}_B(v) = \begin{cases} 0, & i = 1 \\ \frac{y_i - y_{i-1}}{x_i - x_{i-1}}, & i = 2, \dots, n \end{cases} \quad (10)$$

For more details, see, See; Burden and Faires (2011) and Levy (2012).

**Theorem 4.** with one-sided 0 at the endpoints of the data set, the backward difference approach is

$$\hat{F}_B(v) = \begin{cases} 0, & i = 1 \\ \frac{i-1}{n}, & i = 2, \dots, n \end{cases}$$

*Proof:*
By noting that $y_i - y_{i-1} = \frac{i-1}{n}(x_i - x_{i-1}), \quad i = 2, \dots, n$.

This shows that the backward difference approach for left MAD function is just the empirical distribution function $\frac{i-1}{n}, i = 1, \ldots, n$ and has equal jumping value $1/n$.

## 6.3 Centre difference approach

More accurate scheme can be derived using Taylor series

$$g(x+h) = g(x) + \acute{g}(x)h + \frac{\acute{g}(x)}{2}h^2 + \frac{g^{(3)}(\xi_1)}{6}h^3$$

and

$$g(x-h) = g(x) - \acute{g}(x)h + \frac{\acute{g}(x)}{2}h^2 - \frac{g^{(3)}(\xi_2)}{6}h^3$$

By subtracting, the second order approximation ($O(h^2)$) of the first derivative is

$$\acute{g}(x) = \frac{g(x+h) - g(x-h)}{2h} - \frac{h^2}{6}g^{(3)}(\xi) = \frac{g(x+h) - g(x-h)}{2h} + O(h^2),$$

and $\xi \in (x-h, x+h)$, see; Burden and Faires (2011) and Levy (2012).

With two-sided $1/n$ and $(n-1)/n$ at the endpoints of the data set, an estimate of $F_X(v)$ can be approximated by Taylor series expansion as

$$\hat{F}_C(v) = \begin{cases} \frac{y_2 - y_1}{x_2 - x_1}, & i = 1 \\ \frac{y_{i+1} - y_{i-1}}{x_{i+1} - x_{i-1}}, & i = 2, \ldots, n-1 \\ \frac{y_n - y_{n-1}}{x_n - x_{n-1}}, & i = n \end{cases} \tag{11}$$

**Theorem 5.** with two-sided $1/n$ and $(n-1)/n$ at the endpoints of the data set, the centre difference approach is

$$\hat{F}_C(v) = \begin{cases} \frac{1}{n}, & i = 1 \\ \frac{i}{n} - \frac{1}{n}\frac{(x_i - x_{i-1})}{(x_{i+1} - x_{i-1})}, & i = 2, \ldots, n-1 \\ \frac{n-1}{n}, & i = n \end{cases}$$

*Proof.* By noting that

$$y_{i+1} - y_{i-1} = \frac{i}{n}(x_{i+1} - x_{i-1}) - \frac{1}{n}(x_i - x_{i-1})$$

This shows that the centre difference approach has not equal jumping function as in forward and backward approaches, but it is jumping by unequal quantity that depends on the ratio $(x_i - x_{i-1})/(x_{i+1} - x_{i-1})$ and $n$. Note that $\frac{x_i - x_{i-1}}{n(x_{i+1} - x_{i-1})} \ll 1$ and tends to 0 for $n \to \infty$. Also, note that for $n = 2, \ldots, n-1$, we have

$$\left|\hat{F}_C(v) - \hat{F}_n(v)\right| = \left|\frac{i}{n} - \frac{1}{n}\frac{(x_i - x_{i-1})}{(x_{i+1} - x_{i-1})} - \frac{i}{n}\right| = \frac{1}{n}\frac{(x_i - x_{i-1})}{(x_{i+1} - x_{i-1})} < \frac{1}{n}$$

$\hat{F}_C(v)$ is strongly uniformly consistent as $n \to \infty$; see, Serfling (1980).

## 6.4 FC-Hermite approach

With one-sided difference at the endpoints of the data set, an accurate estimate of $F_X(v)$ can be proposed by what is known as FC-Hermite approach or Hermite spline interpolation from Fritsch and Carlson (1980) and "splinefun" given in stats-package R-software (2021) as

$$\hat{F}_{FCH}(v) = \begin{cases} \dfrac{y_2 - y_1}{x_2 - x_1}, & i = 1 \\ 0.5\left(\dfrac{y_{i+1} - y_i}{x_{i+1} - x_i} + \dfrac{y_i - y_{i-1}}{x_i - x_{i-1}}\right), & i = 2, \dots, n-1 \\ \dfrac{y_n - y_{n-1}}{x_n - x_{n-1}}, & i = n \end{cases} \quad (12)$$

It can be noted that this approach just combines the forward and backward approaches by using three data points at $i+1$, $i$ and $i-1$. The FC-Hermite approach can be rewritten in a very simple form as

$$\hat{F}_{FCH}(v) = \begin{cases} \dfrac{1}{n}, & i = 1 \\ \dfrac{2i - 1}{2n}, & i = 2, \dots, n-1 \\ \dfrac{n-1}{n}, & i = n \end{cases}$$

This approach has equal jumping value $1/n$ except for first and last values and related to Hazen (1914) plotting position. Also, note that for $n = 2, \dots, n-1$, we have

$$\left|\hat{F}_{FCH}(v) - \hat{F}_n(v)\right| = \left|\dfrac{i}{n} - \dfrac{0.5}{n} - \dfrac{i}{n}\right| = \dfrac{1}{2n} < \dfrac{1}{n}$$

$\hat{F}_{FCH}(v)$ is strongly uniformly consistent as $n \to \infty$; see, Serfling (1980).

## 6.5 Forward-backward-center approach

It might be very useful to use the combination of more than one approach to increase accuracy of distribution function estimation. Another proposed estimate for $F_X(v)$ can be obtained by combining forward, backward and centre approaches as

$$\hat{F}_{OBC}(v) = \begin{cases} \dfrac{y_2 - y_1}{x_2 - x_1}, & i = 1 \\ \dfrac{1}{3}\left[\dfrac{y_i - y_{i-1}}{x_i - x_{i-1}} + \dfrac{y_{i+1} - y_{i-1}}{x_{i+1} - x_{i-1}} + \dfrac{y_{i+1} - y_i}{x_{i+1} - x_i}\right], & i = 2, \dots, n-1 \\ \dfrac{y_n - y_{n-1}}{x_n - x_{n-1}} & i = n \end{cases} \quad (13)$$

This can be rewritten as

$$\hat{F}_{OBC}(v) = \begin{cases} \dfrac{1}{n}, & i = 1 \\ \dfrac{3i - 1}{3n} - \dfrac{1}{3n}\dfrac{(x_i - x_{i-1})}{(x_{i+1} - x_{i-1})}, & i = 2, \dots, n-1 \\ \dfrac{n-1}{n} & i = n \end{cases}$$

This approach also uses three data points at $i-1, i,$ and $i+1$ and has advantage of having non equal jumping values. Also, note that for $n = 2, \ldots, n-1$, we have

$$\left|\hat{F}_{OBC}(v) - \hat{F}_n(v)\right| = \left|\frac{i}{n} - \frac{1}{3n} - \frac{1}{3n}\frac{(x_i - x_{i-1})}{(x_{i+1} - x_{i-1})} - \frac{i}{n}\right| = \frac{1}{3n}\left[1 - \frac{(x_i - x_{i-1})}{(x_{i+1} - x_{i-1})}\right] < \frac{1}{n}$$

$\hat{F}_{OBC}(v)$ is strongly uniformly consistent as $n \to \infty$; see, Serfling (1980).

## 6.6 Richardson extrapolation approach

When applying lower order formulas, Richardson's extrapolation is employed to achieve high-accuracy results. see, Levy (2012). As pointed out in Burden and Faires (2011), Richardson extrapolation is significantly more effective with even power than when all powers of $h$ are used because the averaging process creates results with errors $O(h^2)$, $O(h^4)$ and $O(h^6)$, …, with essentially no increase in computation, over the results with errors, $O(h), O(h^2), O(h^3),\ldots$; see, Burden and Faires (2011).

**Theorem 6.** An improved approximation for distribution function estimation based on Richardson's extrapolation is

$$\hat{F}_R(v) = \acute{g}(x) = \frac{2^{2p_1}G(h) - G(2h)}{2^{2p_1} - 1} + O(h^{2p_2}) \tag{14}$$

*Proof.* Assume that $g(h)$ be a difference formula with step-size $h$, approximating $\acute{g}(x)$ as

$$G(h) = \acute{g}(x) + a_1 h^{2p_1} + a_2 h^{2p_2} + a_3 h^{2p_3} + \cdots$$

Note $p_1 < p_2 < p_3, \ldots$ and $a_i$ are constants. Therefore,

$$\acute{g}(x) = G(h) - a_1 h^{2p_1} - a_2 h^{2p_2} - a_3 h^{2p_3} - \cdots = G(h) + O(h^{2p_1})$$

Hence, if we consider $G(h)$ as an approximation to $\acute{g}(x)$, the error is $O(h^{p_1})$. Sure, if $h$ tends to zero, $G(h) \to \acute{g}(x)$. If $G(h)$ is computed for step-size $2h$, then

$$G(2h) = \acute{g}(x) + a_1 2^{2p_1} h^{2p_1} + a_2 2^{2p_2} h^{2p_2} + a_3 2^{2p_3} h^{2p_3} + \cdots$$

Multiply $G(h)$ by $2^{2p_1}$ and subtract from $G(2h)$ we obtain

$$\acute{g}(x) = \frac{2^{2p_1}G(h) - G(2h)}{2^{2p_1} - 1} + [a_2 2^{2p_1} - a_2 2^{2p_2}]h^{2p_1} + \cdots = \frac{2^{2p_1}G(h) - G(2h)}{2^{2p_1} - 1} + O(h^{2p_2})$$

Hence,

$$\acute{g}(x) \approx \frac{2^{2p_1}G(h) - G(2h)}{2^{2p_1} - 1} \tag{15}$$

This new approximation is of order $O(h^{2p_2})$.

The estimate $\hat{F}_R(v)$ can be simply obtained from R software (package 'pracma'; see, Borchers (2021)) using function "numdiff (f=function, x)". But for some cases especially large $n$, $\hat{F}_R(v)$ is not always monotone nondecreasing. To adjust $\hat{F}_R(v)$ to monotone nondecreasing it may use bounded isotonic regression method suggested by Barlow et al. (1972) and Balabdaoui et al. (2009) as follows. The bounded isotonic regression problem is explained as: let $\hat{F}_R(v) = \acute{g}(x)$ be a function of some quantity at the $v = x_i$, with true mean function $g°$. The objective is to estimate $g°$ using least squares by minimizing

$$L(a) = \sum_{i=1}^{n} w_i(y_i - a_i)^2$$

over the class of vectors $a$ such that $a_1 \leq \cdots \leq a_n$ and $w_i \geq 0$ are given weight. The solution is given in Barlow et al. (1972) and Balabdaoui et al. (2009) by min-max formula as

$$\dot{a}_i = \max_{s \leq i} \min_{t \geq i} Av\{(s, \ldots, t)\}$$

Where $Av\{(s, \ldots, t)\} = \sum_{i=s}^{t} y_i w_i / \sum_{i=s}^{t} w_i$; see, Barlow et al. (1972) and Balabdaoui et al. (2009). Van Eeden (1957a, 1957b) incorporated known bounds on the regression function to estimate $L(a)$ under the condition

$$a_L \leq a_i \leq a_U$$

$a_L$ is lower bound and $a_U$ is upper bound. This problem has been solved using a suitable adjustment of the pool-adjacent-violators algorithm, see, Barlow et al. (1972) and Balabdaoui et al. (2009). The computational depends on a suitable functional $M$ defined on the sets $A \subseteq \{1, \ldots, n\}$. This functional for the bounded monotone regression is

$$M(A) = \left(Av(A) \vee \max_A a_L\right) \wedge \min_A a_U$$

Where $\min_A v = \min_{i \in A} v_i$ and $\max_A v = \max_{i \in A} v_i$. The PAVA algorthim has been used to find this solution and has been implemented in the R package OrdMonReg (Balabdaoui et al., 2009) under the function BoundedIsoMean; see, Balabdaoui et al. (2011). This function can produce an estimate that is bounded by $a_L = 0$, $a_U = 1$ and monotone nondecreasing $\hat{F}_{Ra}(v)$ (adjusted Richardson extrapolation). $\hat{F}_{Ra}(v)$ is estimated via function BoundedIsoMean ($y = \hat{F}_R(v)$, $w = 1/n$, a = 0, b = 1) in OrdMonReg package in R software.

**Theorem 7.** The Richardson extrapolation estimator $\hat{F}_{Ra}(v)$ is strongly uniformly consistent

$$\sup_v |\hat{F}_{Ra}(v) - F_X(x)| \to 0 \quad w.p.1$$

*Proof.* Let $F_n(v)$ is the empirical distribution function. It may write

$$|\hat{F}_{Ra}(v) - F_X(v)| \leq |\hat{F}_{Ra}(v) - F_n(v)| + |F_n(v) - F_X(v)|$$

It is we known from Serfling (1980) that

$$\sup_v |\hat{F}_n(v) - F_X(x)| \to 0 \quad w.p.1$$

From (15) it can see that

$$\sup_v |\hat{F}_{Ra}(v) - F_n(v)| \leq \frac{1}{n}$$

tends to 0 when $n \to \infty$. Therefore,

$$\sup_v |\hat{F}_{Ra}(v) - F_X(x)| \to 0 \quad w.p.1$$

**Theorem 8.** The Richardson extrapolation estimator $\hat{F}_{Ra}(v)$ has an asymptotic normal distribution as

$$\sqrt{n}\left(\hat{F}_{Ra}(v) - F_X(v)\right) \to^d N(F_X(v)(1 - F(v)))$$

*Proof.* Let $F_n(v)$ is the empirical distribution function and $F_X(v)$ is the true function. It is well known that; see, Serfling (1980).

$$\sqrt{n}\left(\hat{F}_n(v) - F_X(v)\right) \to^d N(0, F(v)(1 - F(v)))$$

From (15) it can see that

$$\sqrt{n}\left(\hat{F}_{Ra}(v) - F_n(v)\right) \leq \frac{1}{\sqrt{n}}$$

As $n \to \infty$,

$$\sqrt{n}\left(\hat{F}_{Ra}(v) - F_X(v)\right) \to^d N(0, F_X(v)(1 - F(v)))$$

The author has R program to compute adjusted Richardson extrapolation from data upon request.

## 7 Simulation

Simulation study is conducted to evaluate the performance of proposed approaches. Five mixture normal distributions that used in Xue and Wang (2010) are implemented to compare the proposed approaches results with their results. These distributions are Gaussian distribution (G), distribution no.3 (strongly skewed distribution (SS)), distribution no. 5 (outlier (OU)), distribution no.7 (separated bimodal distribution (SB)), distribution no. 14 (smooth comb (SC)). These distributions cover a wide range of shapes and are given in Table 2, for more details; see, Marron and Wand (1992). From every distribution, 1000 simulated samples of sizes 20, 50 and 200 are generated, respectively. The estimators $\hat{F}_n^*$ (empirical), $\hat{F}_O$ (forward), $\hat{F}_B$ (backward), $\hat{F}_C$ (centre), $\hat{F}_{FCH}$ (FC-Hermite), $\hat{F}_{OBC}$ (forward-backward-centre) and $\hat{F}_{Ra}$ adjusted Richardson extrapolation are computed. It should be noted that these estimators are not smoothed, and they will be compared with empirical ($\hat{F}_n^*$) and smoothed estimators' linear spline (PS1), cubic spline (PS3), constrained linear spline (CPS1), and constrained cubic spline (CPS3) given in Xue and Wang (2010) in terms of averaged squared errors (ASE) that is defined as

$$ASE_{\hat{F}} = \frac{1}{n}\sum_{i=1}^{n}\left[\hat{F}_X(v_i) - F_X(v_i)\right]^2$$

Table 2 distribution functions used in the simulation study

| Name | Distribution |
|---|---|
| Standard normal distribution (G) | $N(0,1)$ |
| Strongly skewed distribution #3 (SS) | $\sum_{l=0}^{7}\frac{1}{8}N\left(3\left(\left(\frac{2}{3}\right)^l - 1\right), \left(\frac{2}{3}\right)^{2l}\right)$ |
| Outlier distribution #5 (OU) | $\frac{1}{10}N(0,1) + \frac{9}{10}N\left(0, \left(\frac{1}{10}\right)^2\right)$ |
| Separated bimodal distribution #7 (SB) | $\frac{1}{2}N\left(-\frac{3}{2}, \left(\frac{1}{2}\right)^2\right) + \frac{1}{2}N\left(\frac{3}{2}, \left(\frac{1}{2}\right)^2\right)$ |
| Smooth comb #14 (SC) | $\sum_{l=0}^{5}\left(\frac{2^{5-l}}{63}\right)N\left(\frac{65 - 96/2^l}{21}, \left(\frac{32/63}{2^l}\right)^2\right)$ |

The results of the simulation study are given in Table 3 that illustrates that:

- The ASE decreases for all estimators with increasing $n$,
- The estimators $\hat{F}_O$, $\hat{F}_B$, and $\hat{F}_C$ have almost ASE as $\hat{F}_n^*$ and this is logic where all of them are some types of general class of empirical function; see Cunnane (1978) and Hosking and Wallis (1995),
- The estimator $\hat{F}_M$ has improved ASE classical empirical estimators, for example, if distribution is normal and sample size is 20, there is improvement about %5 in ASE over $\hat{F}_n^*$, $\hat{F}_O$, $\hat{F}_B$ (8.48/8.09),
- The estimator $\hat{F}_H$ is surprised as it is very simple and has a very good improvement in terms of ASE. In all cases, there is an improvement about 10% in ASE over $\hat{F}_n^*$, $\hat{F}_O$, $\hat{F}_B$, about 4% over $\hat{F}_C$, and less improvement about 2% over $\hat{F}_{Ra}$,
- The estimator $\hat{F}_{Ra}$ has a major improvement about 12% over $\hat{F}_n^*$, $\hat{F}_O$, $\hat{F}_B$, medium improvement about 6% over $\hat{F}_C$, and small improvement about 2% over $\hat{F}_H$. The $\hat{F}_{Ra}$ is very comparable to two spline smooth unconstrained estimators PS1 and PS3 in terms of ASE,
- With respect to two monotone nondecreasing constrained splines (CPS1 and CPS3), $\hat{F}_{Ra}$ has less ASE in all studied distributions except normal and $\hat{F}_H$ has ASE almost as same as CPS1 and CPS3,
- For large $n$ as 200, the performance of all estimators is comparable in terms of ASE.

Table 3 averaged squares errors (ASE) of all estimators ($\times 10^3$)

|    | $n$ | PS1* | PS3* | CPS1* | CPS3* | $\hat{F}_n^*$ | $\hat{F}_O$ | $\hat{F}_B$ | $\hat{F}_C$ | $\hat{F}_M$ | $\hat{F}_H$ | $\hat{F}_{Ra}$ |
|---|---|---|---|---|---|---|---|---|---|---|---|---|
| G  | 20  | 6.97 | 6.86 | 8.08 | 7.17 | 8.53 | 8.48 | 8.56 | 8.70 | 8.09 | 7.84 | 7.42 |
|    | 50  | 2.87 | 2.76 | 3.13 | 2.97 | 3.29 | 3.31 | 3.30 | 3.35 | 3.24 | 3.19 | 2.99 |
|    | 200 | 0.82 | 0.75 | 0.83 | 0.78 | 0.84 | 0.84 | 0.83 | 0.83 | 0.82 | 0.81 | 0.77 |
| SS | 20  | 8.07 | 7.36 | 8.70 | 7.89 | 9.02 | 8.58 | 8.66 | 8.89 | 8.21 | 7.90 | 7.24 |
|    | 50  | 3.38 | 2.98 | 3.33 | 3.25 | 3.51 | 3.42 | 3.32 | 3.42 | 3.30 | 3.26 | 3.17 |
|    | 200 | 0.84 | 0.79 | 0.83 | 0.82 | 0.84 | 0.84 | 0.84 | 0.83 | 0.83 | 0.83 | 0.83 |
| OU | 20  | 8.25 | 8.10 | 8.47 | 9.20 | 8.71 | 8.52 | 8.47 | 8.80 | 8.16 | 7.92 | 7.92 |
|    | 50  | 3.38 | 3.33 | 3.41 | 3.38 | 3.46 | 3.32 | 3.32 | 3.38 | 3.27 | 3.22 | 3.22 |
|    | 200 | 0.78 | 0.77 | 0.79 | 0.82 | 0.81 | 0.78 | 0.78 | 0.79 | 0.78 | 0.77 | 0.77 |
| SB | 20  | 8.07 | 7.86 | 8.39 | 8.04 | 8.56 | 8.93 | 7.82 | 8.60 | 8.00 | 7.73 | 7.01 |
|    | 50  | 3.19 | 3.12 | 3.20 | 3.14 | 3.32 | 3.37 | 3.15 | 3.31 | 3.20 | 3.16 | 2.92 |
|    | 200 | 0.79 | 0.77 | 0.79 | 0.78 | 0.81 | 0.84 | 0.83 | 0.84 | 0.83 | 0.82 | 0.79 |
| SC | 20  | 8.20 | 7.80 | 8.28 | 7.98 | 8.56 | 8.62 | 8.48 | 8.82 | 8.14 | 7.85 | 7.18 |
|    | 50  | 3.23 | 3.19 | 3.27 | 3.25 | 3.36 | 3.34 | 3.39 | 3.43 | 3.31 | 3.26 | 3.25 |
|    | 200 | 0.81 | 0.79 | 0.81 | 0.83 | 0.82 | 0.82 | 0.84 | 0.84 | 0.83 | 0.80 | 0.82 |

*The results in these columns from Xue and Wang (2010), G: standard normal, SS: strongly skewed, OU: outliers, SB: separated bimodal, SC: smooth comb.

## 8 Application

In ecotoxicology, lognormal and loglogistic distribution are applied to fit a data. A low percentile 5% is of great interest where the hazardous concentration 5% (HC5) is explained as the value of pollutant concentration protecting 95% of the species; see Posthuma et al. (2010). There is a data set "endosulfant" in R software package "fitdistplus"; see, Delignette-Muller and Dutang (2014, 2021). This data includes acute toxicity values (ATV) for the organochlorine pesticide "endosulfan" (geometric mean of LC50 ou EC50 values in $ug.L^{-1}$), tested on Australian and non-Australian laboratory-species; see, Hose and Van den Brink (2004). Figure 4 displays the MAD plot and Cullen and Frey graph; see, Cullen and Frey (1999). MAD plot shows a very weak right wideness and very weak wideness in the left side. The distribution is very strong right skewed ($k = 0.26$) and has a very long right tail $\hat{T}_R = 6.707$, while very short left tail $\hat{T}_L = 0.266$. The skewness based on tails is 6.441 (very strong). Cullen and Frey graph is an indicative graph where it shows the relationship between Pearson' skewness squared and kurtosis. For given data, the skewness is 5.076 and kurtosis is 30.728 that may suggest a lognormal distribution as a good candidate to fit the data. Moreover, Muller and Dutang (2014, 2021) used lognormal, loglogistic, Pareto and Burr III distributions to fit suitable distribution for ATV data.

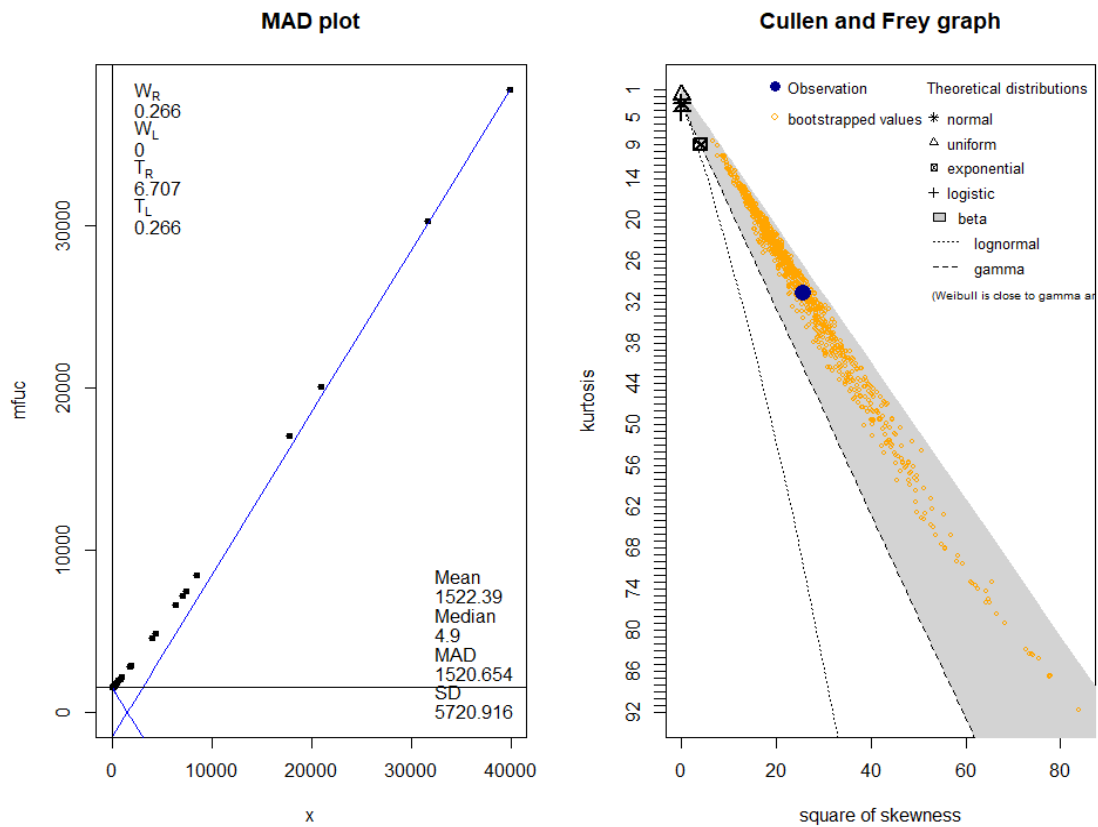

Figure 4 MAD plot and Cullen and Frey graph for acute toxicity values data (ATV)

The proposed adjusted Richardson approximation ($\hat{F}_{Ra}(v)$) is used to estimate nonparametric distribution function of ATV data. Also, 99% pointwise confidence intervals based on normal approximation for $\hat{F}_{Ra}(v)$ are obtained. In Figure 5 the estimated distribution function and 99% confidence interval are plotted along with the estimated parametric distribution functions; for more details about this estimation; see, Muller and Dutang (2021). For lognormal distribution it has a good fit in the right tail while bad fitting at left tail especially due to a very high probability at left tail. The loglogistic does not fit from both tails. The two-parameter Pareto and three-parameter distributions have better fitting in left tail while bad fitting in the right tail. As concluded by Muller and Dutang (2021), none of the four distributions correctly described the right tail observed in ATV data, but the left tail seems to better describe by Burr III distribution; see also, Hose and Van den Brink (2004). They estimated the HC5 value using Burr III distribution as 0.294 while the HC5 from the data is 0.20. The Richardson approximation for HC5 is computed using interpolation as 0.242 while the estimation of HC5 using empirical distribution function is 0.161 (forward approach).

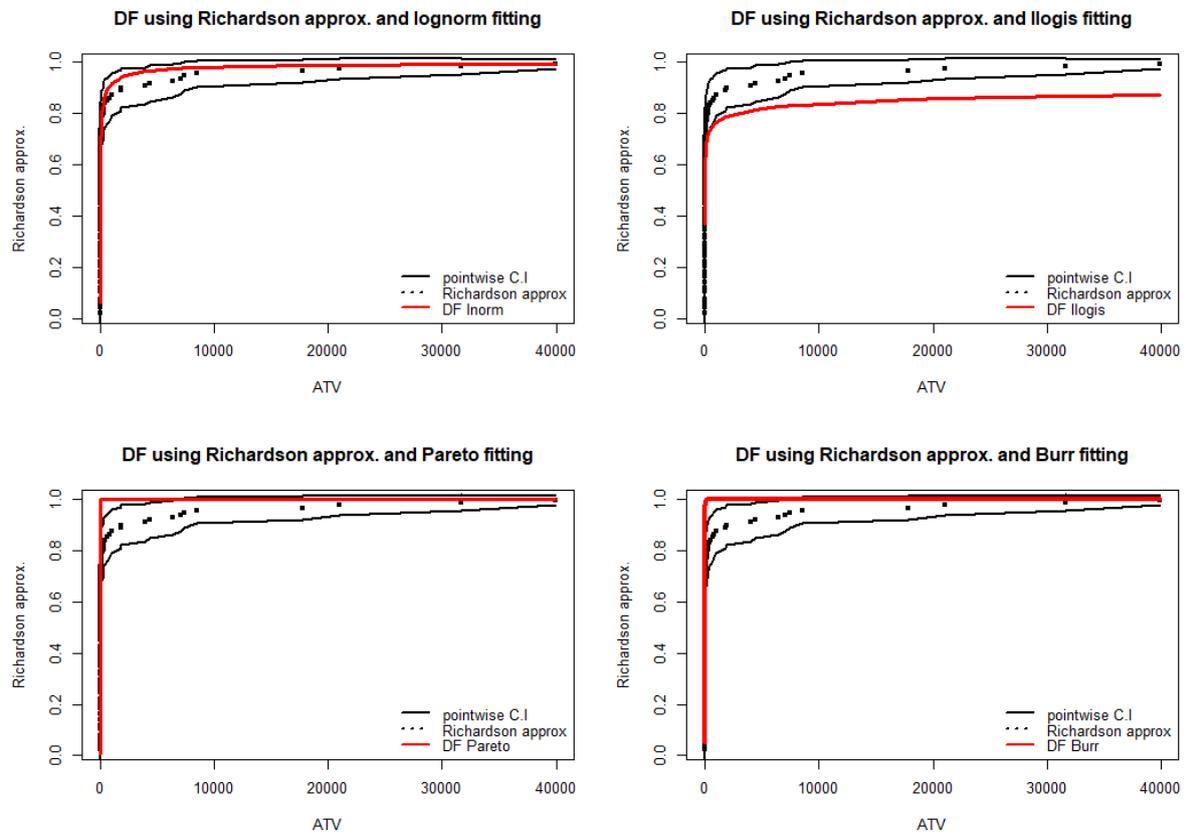

Figure 5 plots of the estimated distribution function using Richardson extrapolation approximation with its 99% pointwise confidence intervals, and the estimated distribution functions from lognormal, loglogistic, Pareto and Burr III models for ATV data.

## 9 Conclusion

The usefulness of the mean absolute deviation function is introduced in two directions. Firstly, it is used to explore the pattern and the structure in the data graphically through the wideness and tailedness. The wideness reflected information about how much the mean absolute function is away from the straight lines $(v_i - \mu)$ and $(\mu - v_i)$. This created right, left and overall wideness measures. These measures could reflect skewness in the data and might use to reflect flatness and peakedness in symmetric distributions. The tailedness reflected information about how long the right and left tails in the data via the maximum of right and left mean absolute deviation functions. A measure of skewness is proposed based on the right and left tails.

Secondly, it is used to estimate the distribution function nonparametrically using numerical differentiation such as forward, central and Richardson approaches. Six approaches had been developed in this direction. One forward, one backward, one central, one mix and one FC-Hermite interpolation and one Richardson extrapolation. Because the Richardson approach did not behave well in large samples, the bounded isotonic regression was suggested to obtain the adjusted Richardson approach. Simulation study was implemented using different distributions that reflected different shapes such as bell-shaped, separated-bimodal, strong-skewed, smooth-comb and outliers. Three estimators showed improvement in terms of averaged squared errors over the classical empirical distribution function. The Richardson extrapolation approach had major improvement in terms of average squared errors over classical empirical estimators and had comparable results with smooth approaches such as cubic spline and constrained linear spline.

Furthermore, the Richardson approach applied for real data application and used to estimate the hazardous concentration five percent. The smoothing approaches such as generalized additive model will be considered in future research to be applied on some of these approaches especially Richardson extrapolation approach.